\documentclass[10pt,a4paper]{article}

\usepackage[left=2.5cm, right=2.5cm, top=2.5cm, bottom=2.5cm]{geometry}

\usepackage{hyperref}
\usepackage{settings}
\usepackage{code}
\usepackage[title]{appendix}

\usepackage{array}
\usepackage{pgfplots}
\usepackage{makecell}
\usepackage{relsize}
\usepackage{subcaption}

\usepackage{tikz}
\usetikzlibrary{positioning}

\newcommand{\Rplus}{\protect\hspace{-.1em}\protect\raisebox{.35ex}{\smaller{\smaller\textbf{+}}}}
\newcommand{\Cpp}{\mbox{C\Rplus\Rplus}}

\newcommand{\id}[1]{\text{INDEX}\left(#1\right)}

\newcommand{\timePlot}[4]{
  \addplot+[memoryDefaults] table[x index=#1, y expr=\thisrowno{0}+#2] {#4};
  \addlegendentry{#3}
}

\newcommand{\timePlotBelow}[4]{
  \addplot+[memoryDefaults, node near coord style ={anchor=north, yshift=-5pt, color=black}] table[x index=#1, y expr=\thisrowno{0}+#2] {#4};
  \addlegendentry{#3}
}

\newcommand{\timePlotLeft}[4]{
  \addplot+[memoryDefaults, node near coord style ={anchor=north, xshift=-15pt, yshift=4pt, color=black}] table[x index=#1, y expr=\thisrowno{0}+#2] {#4};
  \addlegendentry{#3}
}

\pgfplotsset{memoryDefaults/.style={
  only marks,
  mark size=4pt,
  mark options={
    line width=1pt}}}

\definecolor{Red}{RGB}{230, 25, 75}
\definecolor{Green}{RGB}{60, 180, 75}
\definecolor{Yellow}{RGB}{255, 225, 25}
\definecolor{Blue}{RGB}{0, 130, 200}
\definecolor{Orange}{RGB}{245, 130, 48}
\definecolor{Purple}{RGB}{145, 30, 180}
\definecolor{Cyan}{RGB}{70, 240, 240}
\definecolor{Magenta}{RGB}{240, 50, 230}
\definecolor{Lime}{RGB}{210, 245, 60}
\definecolor{Pink}{RGB}{250, 190, 190}
\definecolor{Teal}{RGB}{0, 128, 128}
\definecolor{Lavender}{RGB}{230, 190, 255}
\definecolor{Brown}{RGB}{170, 110, 40}
\definecolor{Beige}{RGB}{255, 250, 200}
\definecolor{Maroon}{RGB}{128, 0, 0}
\definecolor{Mint}{RGB}{170, 255, 195}
\definecolor{Olive}{RGB}{128, 128, 0}
\definecolor{Coral}{RGB}{255, 215, 180}
\definecolor{Navy}{RGB}{0, 0, 128}

\pgfplotsset{ %
  cycle list={ %
    Red, mark=*\\%
    Blue, mark=square*\\%
    Orange, mark=triangle*\\%
    Green, mark=diamond*\\%
    Maroon, mark=asterisk\\%
    Purple, mark=pentagon*\\%
    Teal, mark=o\\%
    Brown, mark=10-pointed star\\%
    Magenta, mark=Mercedes star\\%
  },
}

\newcommand{\green}[1]{{\color{Green}#1}}

\title{Assign optimization for algorithmic differentiation reuse index management strategies}
\author{Max Sagebaum, Johannes Blühdorn, Nicolas R. Gauger}
\date{Chair for Scientific Computing \\ University of Kaiserslautern-Landau (RPTU)}

\begin{document}

\maketitle

\begin{abstract}
  The identification of primal variables and adjoint variables is usually done via indices in operator overloading algorithmic differentiation tools. One approach is a linear management scheme, which is easy to implement and supports memory optimization for copy statements.
  An alternative approach performs a reuse of indices, which requires more implementation effort but results in much smaller adjoint vectors. Therefore, the vector mode of algorithmic differentiation scales better with the reuse management scheme.
  In this paper, we present a novel approach that reuses the indices and allows the copy optimization, thus combining the advantages of the two aforementioned schemes.
  The new approach is compared to the known approaches on a simple synthetic test case and a real-world example using the computational fluid dynamics solver SU2.
\end{abstract}

\section{Introduction}

Algorithmic differentiation (AD) refers to techniques for the machine-accurate differentiation of computer codes.
To that end, we interpret a code with inputs $x\in\R^n$ and outputs $y\in\R^m$ as a mathematical function $F\colon\R^n\to\R^m$.
Usually, the code does not only contain floating point computations but also control flow statements.
For each specific input choice $x$, however, the control flow is fixed and we may consider the computations as a composition of elementary operations like additions or multiplications together with, e.\,g., standard math library functions like $\sin$ or $\exp$.
The derivatives of these elementary operations are known and can be implemented explicitly.
Then, AD evaluates the derivative of the full code at the specific input $x$ by applying the chain rule to the sequence of elementary operations.
In the following, we summarize the principles of AD with a focus on the reverse mode and operator overloading implementations.
Comprehensive introductions to AD are given in \cite{grie08, naumann2012art}.

The reverse mode of AD is a specific strategy for the automatic evaluation of the chain rule.
Each variable $v$ of the computer program --- which might be an input, output, or intermediate variable --- is associated with a corresponding adjoint variable $\bar v$.
$\bar v$ stands for the gradient of the summed outputs $y$ weighted according to $\bar y$ and viewed as a function of $v$, that is, $\bar v = \nabla\left(y(v)^T\bar y\right) = \frac{d y}{d v}(v)^T\bar y$. The adjoint variables $\bar y$ associated with the outputs $y$ --- that is, the weight vector --- must be specified by the user.
Particularly, the adjoint variables $\bar x$ associated with the inputs $x$ satisfy the relation
\begin{equation}
  \bar x = \frac{d F}{d x}(x)^T\bar y
  \label{eq.adjRel}
\end{equation}
which is the standard way to describe the reverse mode of AD.
By suitable choices of $\bar y$, various derivatives of the computer program can be computed.
For example, if $m=1$ and $\bar y=1.0$, $\bar x$ is the gradient of $F$ evaluated at $x$.

During a single evaluation of \eqref{eq.adjRel}, the reverse mode of AD avoids the computational cost of setting up the full Jacobian $\frac{d F}{d x}$.
Instead, an evaluation on the statement level is performed that is equivalent to the computation of the matrix-vector product.
Let $w = \phi(v)$ be a statement in the computer code with $\phi\colon \R^k \rightarrow \R$ and $k \in \N$.
$\phi$ might be a single elementary operation or a composition of multiple elementary operations.
For each statement, reverse AD computes the adjoint update as
\begin{equation}
  \bar v_j \aeq  \frac{\d \phi}{\d v_j}(v) \bar w \quad \forall j = 1 \ldots k; \quad \bar w = 0 \eqdot
  \label{eq.genRev}
\end{equation}
The order of statement evaluation is reversed, that is, if we have statements $\phi_i$ with $ i = 1 \ldots s$ for the primal evaluation, we need to evaluate Equation \eqref{eq.genRev} for $i$ running from $s$ to $1$.
In more general terms, the adjoint information is propagated from the outputs to the inputs, which corresponds to the evaluation of the chain rule in reverse order.
Since the reverse run requires information that is only available after the primal evaluation is finished, certain information needs to be stored by AD such that the reversal is possible.
The process of recording this information alongside the primal computation is known as taping. In operator overloading AD tools, the tape is created by exchanging the computational type, like \ic{double}, with a so-called active type, like \ic{codi::RealReverse}. The active type overloads all operators and mathematical functions. In addition to the primal computational logic, the overloads are used to store the information for the reverse AD path on the tape.

Since the reverse propagation is decoupled from the recording, the adjoint variable $\bar v$ corresponding to the primal variable $v$ cannot be stored directly alongside the primal variable. It is an established practice for operator overloading AD tools to store the adjoint variables in a so-called \emph{adjoint vector}. This is done by assigning an index to each variable, $\id{v} \in \N$, and accessing the adjoint via the adjoint vector, that is, $\bar v = adj(\id{v})$.

Currently, there are two major techniques for distributing the indices, linear index management and reuse index management. Linear index management has the advantage that copy statements like $c = a$ do not need to be recorded on the tape, which usually yields a substantial reduction in the memory required by the tape. On the other hand, the size of the adjoint vector with the reuse index management is by magnitudes smaller than the one with the linear index management. This does not only have advantages for the reverse evaluation speed but also provides better memory and runtime scaling for the vector mode of AD, where the memory consumed by adjoints scales with the vector dimension $d \in N$. We are not aware of any previous systematic comparison of index management approaches, therefore a brief summary of the schemes and their distinguished properties is given in Sections \ref{sec.linearManagement} and \ref{sec.reuseManagement}. A detailed review of the two approaches is available in version 2 of this paper \cite{sagebaum2021index}.

The key contribution of this paper is a reference counting strategy tailored to AD such that reuse index managers can also apply the copy optimization. This reduces the size of the adjoint vector further, which improves the memory consumption and speed of the AD evaluation. We present this approach in Section \ref{sec.multiManagement} and compare its properties with the known management schemes.

In Section \ref{sec.results}, the three schemes are compared with respect to memory and runtime in a benchmark based on a numerical solver for a partial differential equation and in the multiphysics suite SU2 \cite{economon2015su2} on a real-world test case.
All code presented in this paper is \Cpp11 code.

\section{Linear index management and copy optimization}
\label{sec.linearManagement}

The linear index management strategy is used in AD tools like dco/c++ \cite{AIB-2016-08} or CppAD \cite{bell2022cppad}. An exemplary implementation is also available in \cite{grie08} in Section 6.1.
Let $s$ be the total number of statements, the basic idea of the strategy is that for each operation $w_i = \phi_i(\ldots)$ a unique index is created, that is, $\id{w_i} \in \N$ with $\id{w_i} \not = \id{w_j}$ for all $i ,j \in \{1, \ldots, s\}$ with $i \not = j$. Therefore, there is a one-to-one relation between adjoint variables and entries in the adjoint vector.  Therefore, the adjoint vector requires $s \cdot \text{MEM(double)}$ bytes of memory. Usually, the implementation of the linear index management is done such that there is a global variable \ic{indexCounter}. Each time a statement is recorded on the tape, this counter is incremented by one and the new value is used as the index.

The copy optimization is used in dco/c++ for linear index management, for which we recapitulate the basic idea here. A copy statement such as
\begin{equation}
  c = a
\end{equation}
with $c, a \in \R$ is reversed as
\begin{equation}
  \bar a \aeq \bar c; \quad \bar c = 0 \eqdot
\end{equation}
Figure 1 illustrates the life cycle of $a$ and $c$, together with the life cycle of the corresponding adjoint variables. As can be seen in the latter, $\bar c$ and $\bar a$ are only used in a linear fashion in all operations, and except as denoted in steps R1 and R3, neither $\bar a$ nor $\bar c$ appear on the right-hand side of an update. Therefore, it is safe to add contributions to $\bar c$ directly to $\bar a$ instead. This is achieved by removing the copy operation from the tape, that is, drop step R3, and identify $\bar c$ with $\bar a$ by assigning INDEX(c) = INDEX(a).
\begin{figure}
\begin{center}
  \begin{tikzpicture}[node distance=1cm]
  \tikzset{
    fitting node/.style={
      inner sep=0pt,
      fill=none,
      draw=none,
      reset transform,
      fit={(\pgf@pathminx,\pgf@pathminy) (\pgf@pathmaxx,\pgf@pathmaxy)}
    },
    reset transform/.code={\pgftransformreset}
  }

  \draw[->] (0,0) node[left]{forward eval.} -- (8,0) node[right, align=left] {finish};
  \draw[->] (0.5,0.5) node[above,align=center,draw](f1){$a$ assigned \\ $a = \ldots$} -- (0.5,0);
  \draw[->] (2.75,0.5) node[above,align=center,draw](f2){$a$ is used} -- (2.75,0);
  \draw[->] (5,0.5) node[above,align=center,draw](f3){$a$ copied to $c$ \\ $c = a$} -- (5,0);
  \draw[->] (7.5,0.5) node[above,align=center,draw](f4){$a$ and $c$ \\ are used} -- (7.5,0);

  \draw[<-] (0,-1.25) node[left]{reverse eval.} -- (8,-1.25) node[right, align=center] { $\bar a$ and $\bar c$ \\ are zero};
  \draw[->] (0.5,-1.75) node[below,align=center,draw](r1){$\bar a$ is used for \\ an update\\ $\ldots \aeq J * \bar a$} -- (0.5,-1.25);
  \draw[->] (2.75,-1.75) node[below,align=center,draw](r2){ $\bar a$ is \\ updated \\ $\bar a \aeq \ldots$} -- (2.75,-1.25);
  \draw[->] (5,-1.75) node[below,align=center,draw](r3){Add updates \\ of $\bar c$ to $\bar a$ \\ $\bar a \aeq \bar c$} -- (5,-1.25);
  \draw[->] (7.5,-1.75) node[below,align=center,draw](r4){ $\bar a$ and $\bar c$ \\ are updated \\ $\bar a \aeq \ldots$ \\ $\bar c \aeq \ldots$} -- (7.5,-1.25);

  \draw[->] (8.75,-0.25) -- (8.75,-0.75);
  
  \node[below = 0.0em of f1, xshift=-0.75em]{F1};
  \node[below = 0.0em of f2, xshift=-0.75em]{F2};
  \node[below = 0.0em of f3, xshift=-0.75em]{F3};
  \node[below = 0.0em of f4, xshift=-0.75em]{F4};
  
  \node[above = 0.0em of r1, xshift=-0.75em]{R1};
  \node[above = 0.0em of r2, xshift=-0.75em]{R2};
  \node[above = 0.0em of r3, xshift=-0.75em]{R3};
  \node[above = 0.0em of r4, xshift=-0.75em]{R4};
  \end{tikzpicture}
\end{center}
\caption{Life cycle of two variable that are copies of each other.}
\label{fig.relationCopy}
\end{figure}

The amount of tape memory saved by the copy optimization depends on the number of copy statements in the application. However, since the frequently applied call-by-value and return-by-value paradigms require copies, there is usually quite a substantial reduction in memory.
An example implementation of a linear index manager can be found in Appendix \ref{sec.linarImplementation}.

\section{Reuse index management}
\label{sec.reuseManagement}

The reuse index management strategy is used in AD tools like ADOL-C \cite{Walther2012Gsw} or  Adept \cite{Hogan2014FRM}. In Section 4.1 in \cite{grie08}, the basic rules concerning the aliasing effect during the distribution of indices are discussed. The key observation is that the lifetime of a variable --- and hence also its index --- can be quite short. At different points during the execution of the program, different variables are stored at the same memory location. It makes sense to reuse indices of overwritten or freed variables so that also different adjoint variables share a single entry in the adjoint vector.

Two adjoint variables can share a location in the adjoint vector when the lifetime of the associated primal variables does not overlap, that is, the first variable is destroyed before the second variable is instantiated. Since we use indices to associate primal variables with adjoint variables, the index of the first variable needs to be freed before the index is reused in the second variable. Proper implementations of the constructors, destructors, and assign operators of the overloaded AD type can ensure this behavior. The application has to ensure that no other operations create a copy of a variable and associated index. Such a copy breaks the uniqueness of the index, which leaves the reuse index manager in a state of undefined behavior.

Commonly, an implementation of the reuse index management creates a list \ic{availableIndices} and adds the indices of overwritten or freed variables to that list. If a new index is required, it is taken from the list. An example implementation of a reuse index manager can be found in Appendix \ref{sec.reuseImplementation}.

\section{Reuse index management and copy optimization}
\label{sec.multiManagement}

To apply the copy optimization strategy to reuse index management, we need to lift some of the restrictions in Section \ref{sec.reuseManagement}. We have to assume that the first variable has $l$ copies, which all hold the same index. Then, two adjoint variables can share a location in the adjoint vector when the lifetime of the associated variables and all their copies does not overlap. This means that the first variable and all $l$ copies need to be destroyed before the lifetime of the second variable begins. This can be ensured by using a reference counting technique for the indices.

A solution offered by the \Cpp\ standard for such a reference counting scheme is the use of smart pointers \cite{christopher1984reference,meyers2014effective}. Instead of storing an index for each variable, e.g. \ic{int index}, it could be replaced with a smart pointer, e.g. \ic{std::shared_ptr<int> indexRef}. Every time a variable is copied, the smart pointer would also be copied, and on a release, the smart pointer is also released. The smart pointer with the last reference would then use a custom \ic{Deleter} to add the freed index to the pool of available indices.

The use of smart pointers has memory and performance drawbacks in our case. Each shared pointer has a size of $16$ bytes since it has to hold the pointer and a pointer to the reference counter. In addition, the memory for the reference counter amounts to at least $12$ bytes, $4$ bytes for the counter and $8$ bytes for the function pointer to the \ic{Deleter}. Performance-wise, each access to the index requires a dereference operation, also when copying or freeing the index, since the reference counter needs to be accessed.

We want to improve on the baseline solution with shared pointers by using the domain-specific knowledge for AD. Instead of allocating a separate reference counter for each index, we allocate one large vector \ic{std::vector<int> useCount} for the reference counting. The index itself is used as the key for the lookup. This has two advantages. First, all variables keep the same size, since they only need to store the index. Second, the lookup for the reference counting is now an array lookup which can benefit more from caching effects. The basic implementation for handling the reference counting with the \ic{useCount} approach is shown in Listing \ref{fig.useCountImpl}.
\vspace{0.5em}
\begin{codeRefInline}{fig.useCountImpl}{multi use manager for indices.}
struct IndexUseCount {
  std::vector<int> useCount;
  
  void setMaximumSize(size_t size) {
    useCount.resize(size);
  }
  
  int count(int i) {
    return useCount[i];
  }
  
  void unuseIndex(int i) {
    useCount[i] -= 1;
  }
  
  void useIndex(int i) {
    useCount[i] += 1; // new use location of the index
  }
};
\end{codeRefInline}
\ic{useIndex} is called for each copy operation and \ic{unuseIndex} is called for each free of an index. \ic{count} can be used to check if the last reference to the index is freed and therefore is safe to be used again.

Table \ref{tab.solutionComp} compares, no reference counting, the shared pointer solution, and the proposed \ic{useCount} solution. The \ic{useCount} vector solution requires less memory than the shared pointer solution and should be faster since it only requires array accesses instead of dereference operations. In comparison with no reference counting, each index requires an additional $4$ bytes in the \ic{useCount} vector and each free and copy operation requires now an array access. For the copy operation, the array access is negligible compared to the memory operations for the creation of a statement on the tape.

\begin{table}
  \caption{Comparison of the shared pointer reference counting with the AD specific \ic{useCount} vector reference counting in contrast to no reference counting.}
  \begin{center}
    \begin{tabular}{lccc}
    \hline
    & no counting & shared pointer & \ic{useCount} vector\\
    \hline
    Memory in each variable & 4 byte & 16 byte & 4 byte \\
    Memory for each index & 0 byte & 12 byte & 4 byte \\
    Index access dereference operations & 0 & 1 (pointer) & 0 \\
    Copy dereference operations & 0 & 1 (pointer) & 1 (array) \\
    Free dereference operations & 0 & 1 (pointer) & 1 (array) \\
    \hline
    \end{tabular}
  \end{center}
  \label{tab.solutionComp}
\end{table}

The class \ic{IndexUseCount} can now be used to implement a new index manager named \ic{MultiUseIndex- Manager}. The full implementation is shown in Appendix \ref{sec.multiUseImplementation}. The same strategy is also implemented in a new index manager in CoDiPack \cite{CoDiPack}. Unlike linear index management, the new index manager is not compatible with C-like memory operations because index copies have to be accompanied by incrementing the respective reference counters. Based on the implementations in Appendices \ref{sec.linarImplementation} to \ref{sec.multiUseImplementation}, we provide a comparison of all three index managers in Table \ref{tab.summary}. It can be seen that the linear index manager has a very low implementation complexity since neither branching nor any array lookups are required. The implementation complexity for the multi use index manager is increased in contrast to the reuse index manager.

\begin{table}
  \begin{center}
  \begin{tabular}{lccc}
    \toprule
    & \ic{LIM} & \ic{RIM} & \ic{MUIM} \\
    \midrule
    Memory per statement (byte) & 0 & 4 & 4 \\
    Memory per index (byte) & 0 & 0 & 4 \\
    Branches for assign & 0 & 3 & 4 \\
    Array lookups for assign & 0 & 1 & 2 \\
    Branches for free & 0 & 2 & 3 \\
    Array lookups for free & 0 & 1 & 2 \\
    Branches for copy & 0 & 4 & 5 \\
    Array lookups for copy & 0 & 1 & 3 \\
    Copy optimization & + & - & + \\
    Special properties & C-mem comp. & - & - \\
    \bottomrule
  \end{tabular}
  \end{center}
  \caption{Summary of properties and implementation details for the three index managers: \ic{LinearIndexManager} (\ic{LIM}), \ic{ReuseIndexManager} (\ic{RIM}), \ic{MultiUseIndexManager} (\ic{MUIM}). Note that this overview does not contain memory consumption due to the adjoint vector.}
  \label{tab.summary}
\end{table}

In the following, we formulate equations for the precise memory consumption due to the different index managers.
Table \protect\ref{tab.summary} gives rise to the definitions in equations \protect\eqref{eq.defStart} to \protect\eqref{eq.defEnd}.
\begin{align}
  s_a &:= \text{number of assign statements} \label{eq.defStart}\\
  s_c &:= \text{number of copy statements}\\
  s_i &:= \text{number of register input statements} \\
  s & := s_a + s_c + s_i \text{ (total number of statements)} \\
  i_\text{max} & := \text{maximum number of simultaneously active AD variables} \\
  d & := \text{vector mode dimension} \\
  m_d & := \text{\ic{sizeof(double)}} = 8 \text{\,bytes}\\
  m_i & := \text{\ic{sizeof(int)}} = 4 \text{\,bytes} \label{eq.defEnd}
\end{align}
The distinction between usual assign statements ($s_a$) and copy statements ($s_c$) is due to the copy optimization for the linear index manager and the new multi use index manager. For reuse index managers, $i_\text{max}$ is equal to the maximum index assigned to an AD value.

With these definitions, the memory consumption of the index managers are given by Equations \protect\eqref{eq.memStart} to \protect\eqref{eq.memEnd}. Note that we show only the memory with respect to the index management. Memory for each statement, argument, or other management memory is not considered.
\begin{align}
  \text{MEM(LIM)} &= m_d * (s_a + s_i) * d \label{eq.memStart} \\
  \text{MEM(RIM)} &= m_d * i_\text{max}(\text{RIM}) * d + m_i * (s_a + s_c) \label{eq.memMid} \\
  \text{MEM(MUIM)} &= m_d * i_\text{max}(\text{MUIM}) * d + m_i * s_a  + m_i * i_\text{max}(\text{MUIM}) \label{eq.memEnd}
\end{align}
The linear index manager (LIM) generates an adjoint vector that has the size of the number of statements minus the number of copy statements due to the copy optimization.
However, it does not generate memory for each distributed index.
The other two index managers generate adjoint vectors that are sized according to the number of AD variables in the program ($i_\text{max}$).
Usually, these are smaller by orders of magnitude compared to the linear index management case.

Unlike tapes using a linear index manager, tapes using a reuse index mangers need to store the left hand side index for each statement. The number of storage locations is almost as large as the number of entries in the adjoint vector of the linear index manager case. Therefore, if \ic{int} is used for storing indices and \ic{double} is used for storing adjoints, memory gains in a non vector mode setting ($d=1$) can only be as large as half the size of the adjoint vector of the linear index manager case.
The new multi use index manager has the advantage that the memory for indices is smaller because statements for copy operations are removed from Equation \eqref{eq.memEnd}.
However, extra memory is required for the multi use of each index. Compared to the reuse index manager, memory is saved with the multi use index manager as soon as $i_{\text{max}}<s_c$. Since $i_{\text{max}}$ is usually quite small, this often holds true in practice.

Whether the memory consumption of a linear or reuse scheme is preferable depends on the use case.
In a program where $i_\text{max}\approx s$, the linear index manager prevails.
On the other hand, if $i_\text{max} \ll s$, then the reuse index manager is preferable.
Real-world applications will most of the time lie in between these extremes, so a clear recommendation cannot be given.

However, the situation changes in the vector mode setting. $d$ is usually chosen either as a multiple of the SIMD vector size or equal to the size of the output variables of the program.
Here, the reuse index managers require a lot less memory than the linear index manager. For the linear index manager, $d$ is multiplied by the number of statements, which is usually quite large.
From Equations \eqref{eq.memMid} and \eqref{eq.memEnd} it would seem that the adjoint vectors of both reuse managers have the same size. However, the new multi use index manager changes the index distribution of the program. The elimination of copy operations on the tape can lead to a reduction in $i_{\text{max}}$, which can improve the scaling in the vector mode even further.

The SU2 results in Table \protect\ref{tab.su2Memory} give an impression of the achievable memory savings. For $d=1$, the adjoint vector requires 13.67\,GB for the linear index manager, 0.64\,GB for the reuse index manager, and 0.52\,GB for the multi use index manager.
As discussed above, these sizes would scale with $d$.

\section{Results}
\label{sec.results}

\subsection{Coupled Burgers' equations}

The coupled Burgers' equations are used for a general comparison of performance values for the different implementations.
Problem setup and discretization are already described in \cite{SaAlGauTOMS2019} and is also used, e.\,g., in \cite{bluehdorn2023Event}. For completeness, we recapitulate the problem formulation here.

The coupled Burgers' equation \cite{biazar2009exact,bahadir2003fully,zhu2010numerical}
\begin{align}
  u_t + uu_x + vu_y &= \frac{1}{R}(u_{xx} + u_{yy}), \\
  v_t + uv_x + vv_y &= \frac{1}{R}(v_{xx} + v_{yy})
\end{align}
is discretized with an upwind finite difference scheme.
The initial and boundary conditions are taken from the exact solution
\begin{align}
  u(x, y, t) &= \frac{x + y - 2xt}{1 - 2t^2} \quad (x,y,t) \in D \times \R,\\
  v(x, y, t) &= \frac{x - y - 2yt}{1 - 2t^2} \quad (x,y,t) \in D \times \R
\end{align}
given in \cite{biazar2009exact}.
The computational domain $D$ is the unit square $D = [0,1] \times [0,1] \subset \R \times \R$.
As far as the differentiation is concerned, we choose the initial solution of the time stepping scheme as input parameters, and as the output parameter we take the norm of the final solution.

For the implementation of the program, all methods are written in such a way that they can be inlined and the first 3 runs are used as warm-up runs. All timing values are averaged over 20 evaluations. In addition, the frequency of the CPU is fixed. This yields very stable time measurements which are run on one node of the Elwetritsch cluster at the University of Kaiserslautern-Landau (RPTU).
The node consists of two Intel Xeon 6126 CPUs with a total of 24 cores and 384 GB of main memory.
We discretize the Burgers' equation on a $601\times 601$ grid and solve it with 32 iterations.
As a compiler, gcc version 9 is used. We remark that similar results are obtained with the Intel and clang compiler.

For the time measurements, two different configurations are tested.
\begin{itemize}
  \item The \emph{multi} test configuration runs the same process on each of the 24 cores.
    This setup simulates a use case where the full node is used for computation and every core uses the memory bandwidth of the socket.
  \item The \emph{single} test configuration runs just one process on the whole node.
    This eliminates the memory bandwidth limitations and provides a better view on the computational performance.
\end{itemize}
Both test configurations are evaluated with the two known index managers, namely \ic{LinearIndexManager} (\ic{LIM}) and \ic{ReuseIndexManager} (\ic{RIM}), as well as with the new  index manager \ic{MultiUseIndexManager} (\ic{MUIM}) presented in this paper. All three are used in a Jacobian taping approach \cite{SaAlGauTOMS2019} as well as in a primal value taping approach \cite{SaAlGa2018OMS}.

Table \ref{tab.burgersMemory} shows the memory consumption for one process of the Burgers test case. \emph{Statement data} and \emph{argument data} show the required memory for the corresponding tape entries. In addition, we measure the size of the \emph{adjoint vector}.
\emph{Total memory} is a combination of the aforementioned three and contains all additional minor memory sources. Since the test case is optimized such that there are nearly no copy operations, the memory values of the new multi use index manager do not improve by a large factor. Nevertheless, the few copy operations have been eliminated, which can be seen in the memory of the argument data. It decreases from $3971.81$\,MB for the reuse index manager to $3963.54$\,MB for the multi use index manager. This is also indicated by a small reduction in the statement data for the multi use index manager with respect to the reuse index manager. In addition, the memory for the argument data is the same for the linear index manager and the multi use index manager.

\begin{table}
  \begin{center}
  \begin{tabular}{lcccc}
    \toprule
    type & adjoint vector & statement data & argument data & total memory\\
    \midrule
    Jacobian LIM & 711.77 MB & \green{88.97} MB & \green{3963.54} MB & 4764.28 MB \\
    Jacobian RIM & \green{16.50} MB & 444.86 MB & 3971.81 MB & 4433.18 MB \\
    Jacobian MUIM & \green{16.50} MB & 441.41 MB & \green{3963.54} MB & \green{4429.72} MB \\
    \midrule
    primal LIM & 711.77 MB & \green{1512.70} MB & \green{1849.80} MB & 4074.26 MB \\
    primal RIM & \green{16.50} MB & 1868.40 MB & 1852.50 MB & 3754.17 MB \\
    primal MUIM & \green{16.50} MB & 1853.93 MB & \green{1849.80} MB & \green{3745.20} MB \\
    \bottomrule
  \end{tabular}
  \end{center}
  \caption{Memory requirement for the Burgers test case.}
  \label{tab.burgersMemory}
\end{table}

Since there are nearly no copy operations, the Burgers test case can be used to analyze the overhead of the reference counting implementation. Figure \ref{fig.burgersRecording} shows the timings for the Jacobian types and primal values types for the recording of the tapes. The multi use index manager is compared to the reuse index manager, the drop in performance for the Jacobian taping approach is 9\% for the single case and 4\% for the multi case. The larger performance drop in the single case might be explained by some caching effects. For more data intense modes such as primal value taping or the multi cases, the cache is already saturated. If the saturation is not present in the reuse index management mode of the Jacobian single case, it might be reached due to the additional data access in the multi use management scheme. For the primal value taping the overhead is about 5\%. The multi load case is the natural one for HPC and therefore it can be concluded that the multi use index manager has only a minimal overhead for the recording performance of the tapes compared to the classical reuse index management.

The timing results for the reversals of the tapes are shown in Figure \ref{fig.burgersReverse}. Since the multi use index manager only changes the logic for the recording of the tape and the Burgers test case contains next to no copy operations, we expect that the results for the reuse index manager and multi use index manager are nearly the same. This is the case for the Jacobian taping and the primal value taping approach. The performance drop for the primal value tapes from the linear index manager to the reuse index managers is against the expected trend. The drop is about 12\% for the single case and 6\% for the multi case. We could not determine the cause from an analysis of the assembler code or profiling results. The only hint in the profiling results is a higher cost of evaluating the function pointers. Why that is the case could not be determined.

As expected, the multi use index manager has a small negative impact on performance during the recording of a tape.
In the reverse interpretation of the tape, no performance gains are expected since the Burgers test does not involve many copy operations.
\tikzstyle{arrowDef}=[arrows=->, line width=1.5pt, color=purple]
\tikzstyle{nodeDef}=[black, font=\footnotesize, align=left]
\begin{figure}
  \centering
  \begin{tikzpicture}
    \begin{axis}[
      height=5.5cm,
      width=0.9\textwidth,
      xlabel={time in sec.},
      axis x discontinuity=crunch,
      xmin=0.5,
      xmax=2.0,
      ytick={-1, -2, -3},
      ymin=-3.6,
      ymax=-0.4,
      yticklabels={LIM, RIM, MUIM},
      legend style={at={(1.,1.)},anchor=south east, column sep=8pt, cells={anchor=west}},
      legend columns=4,
      transpose legend,
      nodes near coords,
      nodes near coords style={yshift=3pt, color=black},
      point meta=rawx
    ]

      \timePlot{1}{+0.15}{Jacobian single record}{burgersCase/resultsGcc_2023_07_17_gcc_91/codi2GccJacobian_1.dat}
      \timePlot{1}{+0.05}{Jacobian multi record}{burgersCase/resultsGcc_2023_07_17_gcc_91/codi2GccJacobian_24.dat}
      \timePlot{1}{-0.05}{primal single record}{burgersCase/resultsGcc_2023_07_17_gcc_91/codi2GccPrimal_1.dat}
      \timePlotBelow{1}{-0.15}{primal multi record}{burgersCase/resultsGcc_2023_07_17_gcc_91/codi2GccPrimal_24.dat}

    \end{axis}
    \node[] at (5.5cm, 3.35cm) (LIM) {};
    \node[] at (5.85cm, 2.1cm) (RIM) {};
    \node[] at (6.75cm, 0.9cm) (MUIM) {};

    \draw[arrowDef]
      (LIM.south west)
        to[out=-105, in=170]
        node [nodeDef, pos=0.5, left, xshift=-0.25em]
      {cost of index reuse} (RIM.north west);
    \draw[arrowDef]
      (RIM.south)
        to[out=-105, in=170]
        node [nodeDef, pos=0.5, left, xshift=-0.5em] {cost of reference counting}
      (MUIM.north west);
  \end{tikzpicture}
  \caption{Burgers test case tape recording.}
  \label{fig.burgersRecording}
\end{figure}
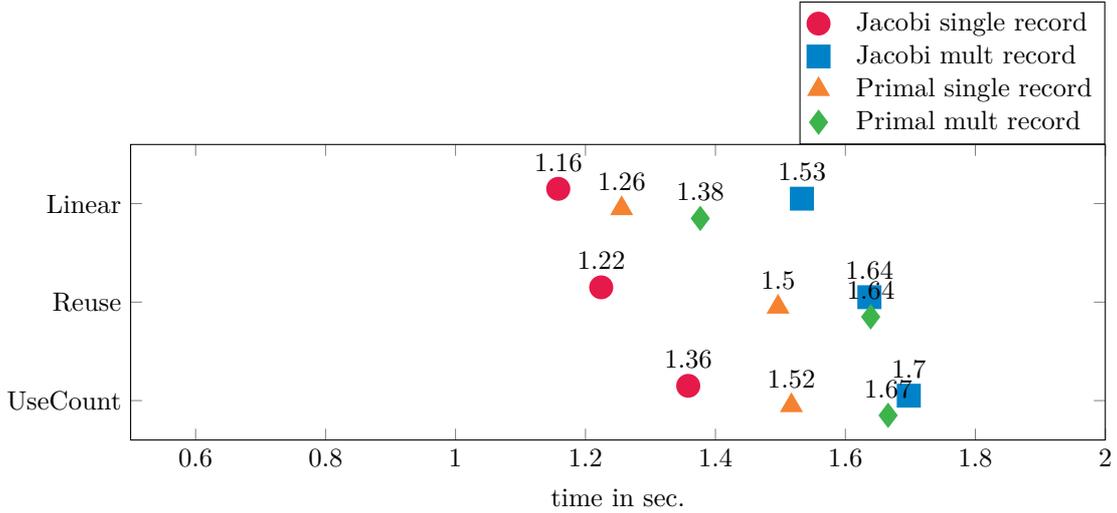
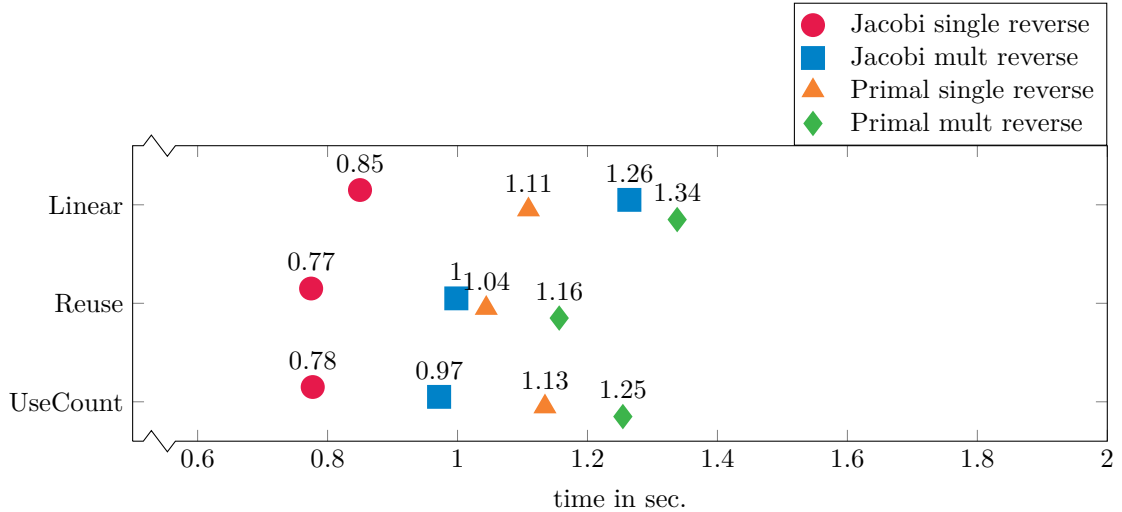
\begin{figure}
  \centering
  \begin{subfigure}{0.5\textwidth}
    \begin{tikzpicture}
      \begin{axis}[
        height=5.5cm,
        width=\textwidth,
        xlabel={time in sec.},
        axis x discontinuity=crunch,
        xmin=0.5,
        xmax=2.0,
        ytick={-1, -2, -3},
        ymin=-3.4,
        ymax=-0.4,
        yticklabels={LIM, RIM, MUIM},
        legend style={at={(1.,1.)},anchor=south east, column sep=8pt, cells={anchor=west}},
        legend columns=4,
        transpose legend,
        nodes near coords,
        nodes near coords style={yshift=3pt, color=black},
        point meta=rawx
      ]

        \timePlot{4}{+0.15}{Jacobian single reverse}{burgersCase/resultsGcc_2023_07_17_gcc_91/codi2GccJacobian_1.dat}
        \timePlot{4}{+0.05}{Jacobian multi reverse}{burgersCase/resultsGcc_2023_07_17_gcc_91/codi2GccJacobian_24.dat}
      \end{axis}
      \node[] at (2.55cm, 3.25cm) (LIM) {};
        \node[] at (2.0cm, 1.8cm) (RIM) {};
        \node[] at (1.85cm, 0.6cm) (MUIM) {};

        \draw[arrowDef] (LIM.south east) to[out=-45, in=30] node [nodeDef, pos=0.5, right, xshift=0.5em] {improvement through \\ smaller adjoint vector} (RIM.north east);
      \draw[arrowDef]
        (RIM.south east)
          to[out=-45, in=40] node [nodeDef, pos=0.5, right] {no improvements expected}
        (MUIM.north east) ;
    \end{tikzpicture}
  \end{subfigure}
  \begin{subfigure}{0.45\textwidth}
    \begin{tikzpicture}
      \begin{axis}[
        height=5.5cm,
        width=\textwidth,
        xlabel={time in sec.},
        axis x discontinuity=crunch,
        xmin=0.5,
        xmax=2.0,
        ytick={-1, -2, -3},
        ymin=-3.6,
        ymax=-0.4,
        yticklabels={},
        legend style={at={(1.,1.)},anchor=south east, column sep=8pt, cells={anchor=west}},
        legend columns=4,
        transpose legend,
        nodes near coords,
        nodes near coords style={yshift=3pt, color=black},
        point meta=rawx
      ]

        \addplot[draw=none] coordinates {(1,1)};
        \addplot[draw=none] coordinates {(1,1)};
        \timePlot{4}{-0.05}{Primal single reverse}{burgersCase/resultsGcc_2023_07_17_gcc_91/codi2GccPrimal_1.dat}
        \timePlotBelow{4}{-0.15}{Primal multi reverse}{burgersCase/resultsGcc_2023_07_17_gcc_91/codi2GccPrimal_24.dat}
        \legend{,,Primal single reverse,Primal multi reverse }
      \end{axis}
    \end{tikzpicture}
  \end{subfigure}

  \caption{Burgers test case tape reversal.}
  \label{fig.burgersReverse}
\end{figure}

\subsection{SU2 Onera M6}

For the second test case, we use the discrete adjoint of SU2 (v7.0.0) \cite{economon2015su2}.
CoDiPack was applied to SU2 some time ago \cite{Albring2015} and the adjoint is based on the primal fixed point formulation
\begin{eqnarray}\label{eq.state_equation}
U = G(U,X) \eqdot
\end{eqnarray}
$X$ represents the design variables and $G(U)$ some (pseudo) time-stepping scheme like the explicit or implicit Euler method.
The fixed-point iteration $U^{n+1} = G(U^n,X)$ is applied until the fixed point $U^*$ is reached.
For the shape optimization with respect to the design $X$, we formulate the minimization problem $min_U \ J(U,X) \ \text{s.t.} \ U = G(U,X)$.
As detailed in \cite{Albring2015}, we solve the minimization problem by fulfilling the KKT conditions \cite{kuhn1951} on the Lagrangian function
\begin{equation}
  \label{eq.lagrangeFunction}
  L(U, X, \bar U) = J(U, X) + (G(U, X) - U)^T \bar U \eqdot
\end{equation}
This requires the solution of the adjoint state equation
\begin{eqnarray}\label{eq.adjstate_equation}
\bar{U} = \left[\frac{\partial J(U^*,X)}{\partial U}\right]^T +  \left[\frac{\partial G(U^*,X)}{\partial U}\right]^T \bar{U},
\end{eqnarray}
which is formulated here as a fixed-point equation.
To solve this equation, we record the tape for the Lagrangian function $L$ with CoDiPack once.
Then, the tape is evaluated again and again until the fixed point $\bar U^*$ is reached.
This solution procedure for the adjoining fixed-point iterations is known as \emph{reverse accumulation} \cite{Christianson1994RAa}. For this method, it is important that the evaluation of the tape is as fast as possible. The recording of the tape plays only a minor role in the performance considerations since a single recording is followed by multiple evaluations.

We consider the viscous flow over the Onera M6 wing as a test case.
The computational mesh consists of $3.6$ million interior elements and the 3D RANS equations are used for the solution.
The test is run on two nodes of the Elwetritsch cluster at the University of Kaiserslautern-Landau (RPTU).
Each node consists of two Intel Xeon 6126 CPUs with a total of 24 cores and 96 GB of main memory each.
In total, the case is run with 192 GB of main memory and 48 cores in an MPI parallel manner.
For the primal computation, one fixed-point iteration step takes about 1.1 seconds and uses 13.73 GB of main memory.

\begin{table}
  \begin{center}
  \begin{tabular}{lcccccrr}
    \toprule
    Type & Adjoint & Statement & Statement & Argument & Tape & Memory & Memory\\
    & vector & entries & data & data & memory & \makecell[c]{change} & \makecell[c]{change}\\
    & (in GB) & & (in GB) & (in GB) & (in GB) & \makecell[c]{LIM} & \makecell[c]{RIM} \\
    \midrule
    Jacobian LIM & 13.666 & 1,834,310,180 & \green{1.753} & \green{46.509} & 61.884 & 0.0\% & -24.9\% \\
    Jacobian RIM & 0.633 & 3,450,714,509 & 16.068 & 65.517 & 82.326 & +33.1\% & 0.0\% \\
    Jacobian MUIM & \green{0.514} & \green{1,750,314,728} & 8.150 & \green{46.509} & \green{55.480} & \green{-10.4\%} & \green{-32.7\%} \\
    \midrule
    Primal LIM & 13.674 & 1,835,297,413 & \green{29.063} & \green{34.456} & 77.194 & 0.0\% & -29.8\% \\
    Primal RIM & 0.633 & 3,452,339,485 & 67.520 & 40.798 & 109.703 & +42.2\% & 0.0\% \\
    Primal MUIM & \green{0.514} & \green{1,751,301,961} & 34.251 & \green{34.456} & \green{70.055} & \green{-9.1\%} & \green{-36.1\%} \\
    \bottomrule
  \end{tabular}
  \end{center}
  \caption{Memory requirement for the SU2 Onera M6 test case. }
  \label{tab.su2Memory}
\end{table}

\begin{figure}
  \centering
  \begin{tikzpicture}
    \begin{axis}[
      height=6.0cm,
      width=0.9\textwidth,
      xlabel={time factor w.r.t. primal},
      ytick={-1, -2, -3},
      yticklabels={LIM, RIM, MUIM},
      ymin=-3.6,
      ymax=-0.4,
      legend style={at={(1.,1.)},anchor=south east, column sep=8pt, cells={anchor=west}},
      legend columns=4,
      transpose legend,
      nodes near coords,
      nodes near coords style={yshift=3pt, color=black},
      point meta=rawx
    ]

      \timePlotLeft{1}{+0.15}{Jacobian Record}{su2OneraCase/results_preAcc_YES/jacobi.dat}
      \timePlotLeft{4}{+0.15}{Jacobian Reversal}{su2OneraCase/results_preAcc_YES/jacobi.dat}
      \timePlotLeft{1}{-0.15}{Primal Record}{su2OneraCase/results_preAcc_YES/primal.dat}
      \timePlotLeft{4}{-0.15}{Primal Reversal}{su2OneraCase/results_preAcc_YES/primal.dat}
    \end{axis}

    \node[] at (4.95cm, 3.85cm) (LIM) {};
    \node[minimum size=1.15em] at (6.55cm, 2.4cm) (RIM) {};
    \node[] at (6.05cm, 1.05cm) (MUIM) {};

    \draw[arrowDef]
      (LIM.south east)
        to[out=-30, in=120, looseness=1.]
        node [nodeDef, pos=0.8, right, xshift=0.5em] {cost of storing copy operations}
      (RIM.north);
    \draw[arrowDef]
      (RIM.south)
        to[out=-90, in=60]
        node [nodeDef, pos=0.5, right, xshift=0.15cm] {gain of copy optimization}
      (MUIM.north east);

    \node[] at (1.75cm, 3.45cm) (LIM) {};
    \node[] at (1.95cm, 2.0cm) (RIM) {};
    \node[] at (1.5cm, 0.6cm) (MUIM) {};

    \draw[arrowDef]
      (LIM.south east)
        to[out=-30, in=30, looseness=1]
        node [nodeDef, pos=0.5, right, xshift=0.5em] {cost of larger tape / \\ gain of reduced \\ adjoint size}
      (RIM.north east);
    \draw[arrowDef]
      (RIM.south east)
        to[out=-30, in=30, looseness=1]
        node [nodeDef, pos=0.5, right, xshift=0.25cm] {reduced tape size / \\ further reduced \\ adjoint size}
      (MUIM.north east);
  \end{tikzpicture}
  \caption{Recording and reversal timing result for the SU2 Onera M6 test case.}
  \label{fig.su2Timing}
\end{figure}

Figure \ref{fig.su2Timing} shows the time factor with respect to the primal time. The results look qualitatively similar for the primal value tapes and Jacobian tapes. A performance improvement of 8\% is seen for the multi use index manager with respect to the reuse index manager. Unlike in the Burgers case, the performance improves for the multi use index manager. This is due to the reduced memory as displayed in Table \ref{tab.su2Memory}, which shows that the increased complexity can be hidden behind the memory bandwidth of the RAM.

The reversal results show the same relative behavior as the recording results, but here, the differences are due to the increased or decreased size of the stored data.
Since the multi use index manager does not store additional statements for the copy operations, it is in general faster than the reuse index manager. For the Jacobian approach, the performance improves by 19\% with respect to the reuse index manager, and the improvement with respect to the linear index manager is 16\%. The primal value taping approach has a performance improvement of 27\% with respect to the reuse index manager and 17\% for the linear index manager.

In contrast to the Burgers case, the memory results in Table \ref{tab.su2Memory} are more interesting. The column of the statement entries can be used to determine $s_a$, $s_c$, and $s_i$ from the values of the different index managers. They are $s_i = 83,995,452$, $s_a = 1,750,314,728$, and $s_c = 1,700,399,781$, which indicates that nearly $50 \%$ of all operations are copy operations. This shows the importance of the new multi use index manager. The reduction in recorded statements shows in the required memory for the tapes. The memory reduction from the reuse index manager to the multi use index manager is about 33\% for the Jacobian taping approach and 42\% for the  primal value taping approach. The memory saving with respect to the linear index manager is only 10\%, which comes from the reduced size of the adjoint vector. The difference in the adjoint vector becomes quite important in the vector mode of AD. For a vector mode of $d = 4$, the Jacobian linear index management approach would use $54.65$\,GB for the adjoint vector and a total of $102.9$\,GB. The Jacobian multi use index management approach would use only $2.06$\,GB for the adjoint vector and a total of $57.02$\,GB, which is a memory reduction of about $45 \%$. The reduction for the Jacobian reuse index management approach is still $33 \%$. A vector mode of $d = 16$ yields a memory reduction of $76 \%$ with respect to the linear index management and $32 \%$ with respect to the reuse index management.

\section{Conclusion}

We reviewed and compared two existing index management schemes. For the reuse index management scheme, we introduced a reference counting extension. It enables the copy optimization technique also for the reuse index management. We presented the basic ingredients for the implementation of the new manager in CoDiPack and analyzed the scheme with respect to its properties and possible effects on the taping process.
The overhead for the reference counting is quite small, which is demonstrated in the Burgers test case. In the bandwidth-limited case, the additional overhead for the reference counting is about 4\%.

The SU2 test case shows the advantages of the new index manager.
The previous best case with regard to memory and runtime was the linear index management strategy in combination with the Jacobian taping approach. The new multi use index manager in combination with the Jacobian taping approach provides an overall performance improvement of about 16\% in combination with a memory reduction of about 10\%. In addition, the scaling for the vector mode of AD is improved by a large margin. The best-case scenario for the AD vector mode in SU2, reuse index management with the Jacobian taping approach, can expect a memory reduction of about $32 \%$.

In general, by eliminating copy operations, our new multi use index management scheme reduces the size of the AD tapes as well as the time to evaluate them. The new scheme might reduce the size of the adjoint vector even further with respect to existing reuse management schemes, thus preserving or improving the scalability with the vector mode of AD.

\begin{appendices}

\section{\ic{LinearIndexManager} implementation}
\label{sec.linarImplementation}

\begin{code}
struct LinearIndexManager {
  int lastIndex;

  LinearIndexManager() : lastIndex(0) {}

  void assignIndex(int& i) {
    lastIndex += 1; // leave out the zero index
    return lastIndex;
  }

  void assignUnusedIndex(int& i) {
    return assignIndex(i);
  }

  void copyIndex(int& lhs, const int& rhs) {
    lhs = rhs;
  }

  void freeIndex(int& i) {
    // empty
  }

  void reset() {
    lastIndex = 0;
  }
};
\end{code}

\section{\ic{ReuseIndexManager} implementation}
\label{sec.reuseImplementation}

\begin{code}
struct ReuseIndexManager {
  int maximumIndex;

  std::vector<int> indices;
  size_t indicesLeft;

  std::vector<int> unusedIndices;
  size_t unusedIndicesLeft;

  static const size_t INDEX_BLOCK_SIZE = 256;

  ReuseIndexManager() :
      maximumIndex(0),
      indices(INDEX_BLOCK_SIZE), indicesLeft(0),
      unusedIndices(INDEX_BLOCK_SIZE), unusedIndicesLeft(0)
  {
    createNewIndices();
  }

  void createNewIndices() {
    // only called if indicesLeft is zero
    for(; unusedIndicesLeft < INDEX_BLOCK_SIZE; unusedIndicesLeft += 1) {
      maximumIndex += 1;
      unusedIndices[unusedIndicesLeft] = maximumIndex;
    }
  }

  void assignUnusedIndex(int& i) {
    freeIndex(i); // force change of index

    if(0 == unusedIndicesLeft) {
      createNewIndices();
    }

    unusedIndicesLeft -= 1;
    i = unusedIndices[unusedIndicesLeft];
  }

  void assignIndex(int& i) {
    if(0 == i) { // leave index in place if not zero
      if(0 == indicesLeft) {
        assignUnusedIndex(i); // fallback to unused indices
      } else {
        indicesLeft -= 1;
        i = indices[indicesLeft];
      }
    }
  }

  void freeIndex(int& i) {
    if(0 != i) {
      if(indicesLeft == indices.size()) {
        indices.resize(indices.size() + INDEX_BLOCK_SIZE);
      }

      indices[indicesLeft] = i;
      indicesLeft += 1;
      i = 0;
    }
  }

  void copyIndex(int& lhs, const int& rhs) {
    if(0 != rhs) { assignIndex(lhs); }
    else         { freeIndex(lhs); }
  }

  void reset() {
    size_t totalSize = indicesLeft + unusedIndicesLeft;
    if(unusedIndices.size() < totalSize) {
      unusedIndices.resize(totalSize);
    }

    std::copy(indices.begin(), indices.begin() + indicesLeft,
              unusedIndices.begin() + unusedIndicesLeft);
    unusedIndicesLeft = totalSize;
    indicesLeft = 0;
  }
};
\end{code}

\section{\ic{MultiUseIndexManager} implementation}
\label{sec.multiUseImplementation}

\begin{code}
struct MultiUseIndexManager {
  int maximumIndex;
  
  std::vector<int> indices;
  size_t indicesLeft;
  
  std::vector<int> unusedIndices;
  size_t unusedIndicesLeft;
  
  IndexUseCount useCount;
  
  static const size_t INDEX_BLOCK_SIZE = 256;
  
  MultiUseIndexManager() :
      maximumIndex(0),
      indices(INDEX_BLOCK_SIZE), indicesLeft(0),
      unusedIndices(INDEX_BLOCK_SIZE), unusedIndicesLeft(0),
      useCount()
  {
    createNewIndices();
  }
  
  void createNewIndices() {
    // only called if indicesLeft is zero
    for(; unusedIndicesLeft < INDEX_BLOCK_SIZE; unusedIndicesLeft += 1) {
      maximumIndex += 1;
      unusedIndices[unusedIndicesLeft] = maximumIndex;
    }
    
    useCount.setMaximumSize(maximumIndex + 1);
  }
  
  void assignUnusedIndex(int& i) {
    
    freeIndex(i); // force change of index
    
    if(0 == unusedIndicesLeft) {
      createNewIndices();
    }
    
    unusedIndicesLeft -= 1;
    i = unusedIndices[unusedIndicesLeft];
    
    useCount.useIndex(i);
  }
  
  void assignIndex(int& i) {
    if(1 == useCount.count(i)) {
      return; // Early out for last use of the index.
    }
    
    freeIndex(i);
      
    if(0 != indicesLeft) {
      // Assign regular index
      indicesLeft -= 1;
      i = indices[indicesLeft];
    } else {
      // Assign unused index
      if(0 == unusedIndicesLeft) {
        createNewIndices();
      }
      
      unusedIndicesLeft -= 1;
      i = unusedIndices[unusedIndicesLeft]
    }
    
    useCount.useIndex(i);
  }
  
  void freeIndex(int& i) {
    if(0 != i) {
    
      useCount.unuseIndex(i);
      if(0 == useCount.count(i)) {
        if(indicesLeft == indices.size()) {
          indices.resize(indices.size() + INDEX_BLOCK_SIZE);
        }
        
        indices[indicesLeft] = i;
        indicesLeft += 1;
        i = 0;
      }
    }
  }
  
  void copyIndex(int& lhs, const int& rhs) {
    if(lhs != rhs) { 
      freeIndex(lhs);
      
      if(0 != rhs) {
        useCount.useIndex(rhs);
        lhs = rhs;
      }
    }
  }

  void reset() {
    size_t totalSize = indicesLeft + unusedIndicesLeft;
    if(unusedIndices.size() < totalSize) {
      unusedIndices.resize(totalSize);
    }
    
    std::copy(indices.begin(), indices.begin() + indicesLeft,
              unusedIndices.begin() + unusedIndicesLeft);
    unusedIndicesLeft = totalSize;
    indicesLeft = 0;
  }
};
\end{code}

\end{appendices}

\bibliographystyle{alphaurl}
\bibliography{citations}

\end{document}